\documentclass[preprint]{aastex}
\usepackage{graphicx}

\tighten
\def\spose#1{\hbox to 0pt{#1\hss}}
\def\ltwig{\mathrel{\spose{\lower 3pt\hbox{$\mathchar"218$}}
     \raise 2.0pt\hbox{$\mathchar"13C$}}}
\def\gtwig{\mathrel{\spose{\lower 3pt\hbox{$\mathchar"218$}}
     \raise 2.0pt\hbox{$\mathchar"13E$}}}
\def\blankline{\par\vskip \baselineskip}
\newcommand{\beq}{\begin{equation}}
\newcommand{\eeq}{\end{equation}}
\newcommand{\beqa}{\begin{eqnarray}}
\newcommand{\eeqa}{\end{eqnarray}}

\newcommand{\momrat}{\eta}

\newcommand{\thetas}{{\theta_s}}
\newcommand{\va}{{v_A}}
\newcommand{\vb}{{v_B}}
\newcommand{\pe}{P_e}
\newcommand{\paz}{P_{A,z}}
\newcommand{\pbz}{P_{B,z}}
\newcommand{\parho}{P_{A,\rho}}
\newcommand{\pbrho}{P_{B,\rho}}
\newcommand{\ps}{P_{s}}

\ifx\pdfoutput\undefined
\else
\fi

\begin{document}
\title{
Asymptotic Opening Angles for Colliding-Wind Bow Shocks:
the Characteristic-Angle Approximation}

\author{Kenneth G. Gayley}
\affil{University of Iowa,
       203 Van Allen Hall,
       Iowa City, IA, 52245}


\begin{abstract}
By considering the advection and interaction of the vector momentum
flux in highly supersonic spherically diverging winds, 
we derive a simple analytic description of the asymptotic 
opening angle of a wind-collision shock cone, in the approximation that the
shocked gas is contained in a cone streaming out along a single 
characteristic opening angle.
Both highly radiative and highly adiabatic limits are treated, and their
comparison is the novel result.
Analytic closed-form
expressions are obtained for the inferred wind momentum ratios as a function of the
observed shock opening angle, allowing the conspicuous shape of the asymptotic
bow shock to be used as a preliminary constraint on more detailed modeling
of the colliding winds.
In the process, we explore from a general perspective
the limitations in applying to the {\it global} shock geometry
the so-called Dyson approximation, which asserts a local balance in the
perpendicular ram pressure across the shock.
\end{abstract}

\section{Introduction}

Wind collisions in close binary systems provide 
a key laboratory for studying astrophysical shocks, including 
X-ray generation, particle acceleration, and dust creation.
They also provide a unique opportunity to study the
attributes of those winds via their hydrodynamical interaction, and it is this latter
advantage that we explore here.
Simulations (e.g., Girard \& Willson 1987; Shore \& Brown 1988;
Eichler \& Usov 2003)
and observations (e.g., Hill, Moffat, \& St-Louis 2002; 
Rauw et al. 2005; Ignace, Bessey, \& Price 2009) 
both indicate that the interaction region forms a large-scale
shock cone whose opening angle is diagnostically 
significant for the parameters of the wind interaction.
Spectral diagnostics in emission or absorption (Luhrs 1997;
St-Louis, Willis, \& Stevens 1993; St-Louis et al. 2005), and the conspicuous
formation of dusty ``pinwheels'' (Tuthill et al. 2008), make it feasible to use the
shock opening angle as a constraint on the wind interaction.

The primary attribute of the winds that control this diagnostic is evidently their 
momentum-flux ratio, subject to the relative efficiency of radiative and adiabatic cooling.
We certainly expect that the shock cone is narrower the greater the contrast between
weak-wind and strong-wind momentum flux, and short radiative
cooling times should also narrow the cone by minimizing the explosiveness of the interaction.
Thus, a simple unified description of how these factors are manifested in the 
asymptotic bow-shock angle is a particularly straightforward way to extract diagnostic
information about these wind attributes.

Although sophisticated modeling, both numerical 
(Girard \& Willson 1987; Shore \& Brown 1988; Comeron \& Kaper 1998)
and analytical (Wilkin 1996; 
Canto, Raga, \& Wilkin 1996; Pilyugin \& Usov 2007), 
has been carried out in both radiative and adiabatic
colliding-wind scenarios, no generalized treatment exists that can extend simple
analytic results for the asymptotic opening angle to both the highly
radiative and highly adiabatic limits.
Certainly, fully detailed results are needed to determine
X-ray flux diagnostics and the complete shock structure, but if we restrict attention
to the asymptotic
opening angle, a conspicuously observable diagnostic,
then these various types of models may be analyzed in a 
simple way that underscores key physical differences in these
limits. 
This is particularly convenient at early stages of analyzing a particular
colliding-wind system, when 
it may not yet be clear which regime to expect.

When radiative cooling is so efficient that gas pressure never plays an important role in
the global shock dynamics, one expects 
the ``thin-shock approximation'' (Girard \& Willson 1987) to provide an adequate treatment,
and in that case
fully analytic results for the asymptotic opening angle
were developed elegantly by Canto, Raga, \& Wilkin (1996; hereafter CRW).
However, this work has not yet been generalized to the opposite limit of slow cooling,
and making that extension for either mixed or unmixed winds is the purpose of this paper.
Neither approximation is without difficulties, as radiating
shocks are subject to instabilities (Stevens, Blondin, \& Pollack 1992)
and adiabatically shocked gas streams will not
remain confined to a thin layer 
owing to explosive expansion (Pittard 2007), but nevertheless the analytic
results give a point of reference that informs 
more detailed and physically realistic simulations.
Having access to simple closed-form results in both limits makes each more useful, 
by virtue of the implied contrasts between them.

\subsection{The standard approach to the steady thin-shock geometry}

When numerical results, rather than analytic forms, 
for the shape of a thin and steady shock are desired,
the standard approach for determining the shock opening angle
involves a detailed integration of the 
shock conditions all along its surface, starting from
the stagnation point along the symmetry axis (when coriolis effects are neglected).
The physical requirement for such an integration is that the shock cone has a
``memory'' of the history of the gas that it is carrying, encoded in the form of
the mass, energy, and momentum fluxes along the shock cone, and the point where
any windstream from one of the stars meets the working surface of the shock 
cone depends on that complete history--- it is not a local attribute of that windstream.
However, a simplification can be applied if it is only the {\it asymptotic} opening
angle that is desired, not the detailed structure.
Then the result should be
expressible in a global form, breaking from
the standard locally integrated numerical approach, and allowing the problem to 
be solved in a single step.

In other words, the opening angle should be solvable by applying purely
global constraints, but so
far that has only been possible for highly radiative thin shocks, as assumed
in CRW.
We suggest that a systematic, if idealized, approximation that allows for
such global constraints follows from a simple extension of the thin-shock
approximation to include adiabatic cooling, but which conserves different
wind properties consistent with the 
presence of explosive gas pressure and the absence of significant radiative cooling.
Then contrasts between the predicted opening angle in the adiabatic
and radiative limits, for 
both mixed and unmixed winds, focus squarely on the effects of adiabaticity, and
guide more complete hydrodynamic studies of these differences.

In the process, we also seek to clarify some potential misconceptions about the relative
importance of the history of
the mass-loading of the shocked region, as opposed
to the local constraints stemming from the collision of two individual windstreams without
consideration of that history.
These issues relate to the concept of centrifugal corrections and the ``Dyson
approximation'' (Dyson 1975), so we begin
our analysis there.

\subsection{The role of centrifugal corrections}

An approximation that has been applied in many
contexts (e.g.,
Luo, McCray, \& Mac Low 1990; Stevens et al. 1992; Antokhin,
Owocki, \& Brown 2004) to highly supersonic bow shocks, 
is requiring a balance between the ram pressure perpendicular to the shock front in the local
colliding windstreams (Dyson 1975).
This amounts to neglecting the momentum requirements of {\it turning} the inertial
flow already moving along the shock front,
so is termed the neglect of ``centrifugal'' corrections.
Explorations into the errors introduced by this approach
(e.g., Mac Low et al. 1991)
typically focus on the behavior near the head of the bow shock and on
the generation of the hardest X-rays, but the approximation
may lose validity farther downstream where the global shock morphology is determined.

It will be argued here that although this approximation is generally valid near the
stagnation point, since there the history of the introduction of
mass into the shocked region is of least importance,
it is not a particularly useful approximation to apply to the 
{\it asymptotic} shape of the
shock, as the latter is more related to globally integrated geometric factors that incorporate
the history of the mass loading of the shock.
As such, the Dyson approximation is of more value for analyzing high-temperature
emissions from near the shock apex, than it is useful for understanding the more
global shock-cone shape.
On the other hand, the global character of our approach cannot produce local emission diagnostics
near the shock apex, so is intended
to be {\it complementary} to the use of the Dyson approximation, 
and applicable only to morphological
observations.

The key global constraint is 
overall conservation of the components of vector momentum flux, tracking
the wind inputs as well as any external exchange of momentum, in
axial wedges conforming to the assumed azimuthal symmetry.
The symmetry axis is taken to be the binary line of centers,
so applies only
on scales where the orbital effects which break that symmetry may be neglected.
As will become more apparent below, tracking the global momentum flux is
substantially different from asserting local perpendicular ram balance,
because the shocked gas carries a momentum flux that must be {\it deflected} moreso
than {\it stopped},
thereby
``soaking up'' locally unbalanced ram pressure, as the shock angle turns from its initially
perpendicular angle to its ultimate asymptotic angle.
This effect represents a type of mass loading, wherein 
the stronger wind piles into previously shocked elements of {\it itself},
and that allows for both global and local violation of the perpendicular ram balance.
This integrated memory of the interaction, not local physics, determines the
asymptotic opening angle of the shock.
As such, this paper explores 
the fundamental {\it conceptual} reason why the Dyson approximation is
not suited to the asymptotic shock geometry, and 
why this is further exacerbated in the presence of adiabaticity.

\section{The formulation of the problem}

In this paper
we consider two highly supersonic spherically diverging constant-velocity
winds colliding along an axis of symmetry in a long-period binary system.
For simplicity, both winds are assumed to have reached their
terminal speed, and we include no dynamical influences, such as radiative
forces or gravity from either star,
nor any orbital effects such as coriolis effects or ellipticity (for a study
of these latter effects, see Lemaster, Stone, \& Gardiner 2007). 
We also assume that the stellar separation is sufficient such that
a normal stagnation point is achieved between the stars-- Luo et al. (1990)
and CRW looked at what happens when one wind impinges directly
on the companion photosphere, but complications ensue and this is not
an entirely solved problem.

The star in the binary 
with the stronger wind momentum flux, if they are not equal, is termed
star A, and the weaker is termed star B.
The goal is to derive the momentum-flux ratio, $\momrat$, of the weak-wind star
B to the strong-wind star A, that is required to support a given observed
asymptotic shock-cone half-angle $\theta_s$, given various assumptions about
the interactions such as if they are radiative versus adiabatic, or mixing versus nonmixing.
The key parameters to track are 
the global fluxes of vector momentum and scalar energy, in a manner
similar to the approach pioneered for 
interstellar medium collisions by Wilkin (1996).
That approach was updated for spherically diverging winds by CRW, and
our main contribution here is the extension to the adiabatic case.
This requires that we track the momentum-flux implications
of the gas pressure and turbulent pressure (which we will treat
as gas pressure), induced by the supersonic collision.

In axial symmetry, the independent domain to understand is an infinitely long
and narrow wedge of
axial angular thicknes $d\phi$ whose vertex is all along the binary line of centers.
This wedge is cut into two pieces by the internal
working surface of the shock cone
between the winds, and 
all the shocked material is assumed to issue out asymptotically
along a single characteristic direction that divides the pieces.
We term this
the ``characteristic-angle'' approximation
to distinguish it from the thin-shock approximation, which 
assumes rapid cooling and explicitly neglects all influences of gas pressure.
Comeron \& Kaper (1998) also relax the requirement of complete radiative
cooling, as applied to wind/ISM interactions, but
do not take the global approach of following the self-consistent gas pressure influences
on the momentum fluxes.

The sources of momentum fluxes in the wedge are the winds of the two stars,
and the combined result is
treated as if it issued from a single point seen from large distance.
As such, we 
do not include the
innovation of CRW of including the angular
momentum flux to track the curving shape of the shock front, 
as we consider the asymptotic domain where the binary system is effectively
resolved into a single point and the angular momentum flux vanishes.
The sole sink of momentum flux is advection through
the outer edge of the wedge at large distances, treated in steady state.

\subsection{The characteristic shock-angle approximation}

Our core assumption, for the purposes of achieving simple closed-form
results, is that the flow of the shocked gas is governed by a single characteristic
angle, $\thetas$, which forms the separatrix of 
the two winds downstream of the interaction region.
Then $\thetas$ is interpreted as the opening half-angle of the asymptotic bow shock, to
within the limitations of the approximation.
This is also the formal limit 
considered by CRW, which they justify by assuming rapid radiative cooling
and lateral confinement of the shocked gas between the two unshocked winds.
When gas pressure is significant, as with adiabatic shocks, such lateral confinement
is not physically realizable, but nevertheless the return of thermal energy to
bulk flow energy, endemic to adiabatic cooling, is assumed to result in the
bulk of the interacting gas exiting the system more or less along a single most
prominent or characteristic direction.
Future hydrodynamic simulations are planned to explore the applicability of this approximation,
but at this stage we view it primarily as a benchmark, intended for systematic
comparison between radiative and adiabatic shocks that support a characteristic
asymptotic opening angle.


\subsection{Solution strategy}

We now lay out our fundamental treatment of the quasi-thin shock approximation,
in terms of conserved fluxes of vector momentum components in the $z$ and $\rho$
directions.
The key addition to the CRW treatment is the inclusion of gas pressure, capable
of generating a new source of momentum flux away from the binary line of centers, as occurs 
in the adiabatic limit.
In axial symmetry, gas pressure does not alter the integrated $z$ component of momentum
flux within each axial wedge under consideration,
because the $z$ direction is globally constant in cylindrical coordinates,
and there are no external forces
in the $z$ direction.
However, gas pressure in the shocked gas can and does increase the flux of the
component of momentum that points perpendicular to the axis, which we denote 
as the $\rho$ direction in cylindrical coordinates.
It is this accounting of the $\rho$ component of the momentum flux that allows us
to track the influences of adiabaticity.

We consider a sphere at very large radius centered on the binary system, 
and track the flux in the colliding winds
of various scalar and momentum quantities through that
sphere.
Owing to the axial symmetry, we may restrict to a thin wedge, of axial angular width $d\phi$
and with its vertex all along the line of centers of the binary, shaped like the section
of an orange.
Considering such identical wedges avoids
cancellation when summing the flux of vectors with $\rho$ components.
We may then apply appropriate global conservation laws, resulting from the given 
flux sources in the 
spherical winds of each
star, and the absence of any external forces (for simplicity
we neglect gravity and radiative forces 
and assume the winds appear at their terminal speeds), to constrain the nature of the
fluxes through the sphere, independently of any wind/wind interaction.
Hence we obtain quantities that must be the same in the actual case as they are in an
imaginary case where the winds do not interact at all, and these quantities supply us
with suitable constraints for deriving the desired expression for
the characteristic shock angle $\thetas$.

\subsection{Basic definitions}

As mentioned above, we define the characteristic angle along which the shocked gas flows
to be $\thetas$, and let $\ps$ be
the scalar momentum flux along that angle (a sum over both shocked
winds).
Purely for convenience we normalize the scalar momentum flux of star A, which
is the mass-loss rate ${\dot M}$ times
the terminal speed $\va$, 
to be $8 \pi$.
As all momentum fluxes in this paper are expressed per angular width 
$d\phi$ of the azimuthal wedge
under consideration,
this implies that they may all be converted to real momentum flux units 
by multiplying them by
${\dot M}_A \va d\phi/8\pi$.

The essential device we use is to consider separately the flux of the component
of momentum in the $z$ direction
(i.e., the component along the binary line of centers), the flux of the component
of momentum in the $\rho$ direction (along the
radius of the wedge), and where possible, the scalar momentum flux (the flux of
the magnitude of the momentum being advected).
Due to the absence of any external forces on the wedge in the $z$ direction, we may assert
equality of the flux of the $z$ component of momentum, between the actual wedge, and an
imaginary wedge where no wind interactions of any kind occur.
This conservation of $z$-component of momentum is written
\beq
\label{zmom}
\ps \cos \thetas \ = \ \paz(\thetas) \ + \ \pbz(\thetas) \ ,
\eeq
where again $\ps$ is the scalar momentum flux of both winds along $\thetas$, and
here $\paz(\thetas)$ and $\pbz(\thetas)$ are
\beq
\paz(\thetas) \ = \ 2 \int_0^\thetas d\theta \ \sin \theta \cos \theta
\ = \ \sin^2 \thetas 
\eeq
and
\beq
\pbz(\thetas) \ = \  -2 \int_\thetas^\pi d\theta \ \sin \theta \cos \theta
\ = \ - \momrat \sin^2 \thetas \ ,
\eeq
which are respectively
the flux of the $z$ component of momentum that is embroiled in the shock
from wind A and from wind B, for $\momrat$ the ratio of the total wind momentum
fluxes in the weak wind (B) to the strong wind (A).

The situation in the $\rho$ direction is altered in a significant way by the potential
presence of gas pressure in the shocked winds.
Due to the azimuthal symmetry,
the pressure may be treated as an equal squeezing force on opposite faces of the wedge,
which when added vectorially, must yield a net flux of momentum in the $\rho$ direction.
Let this source of $\rho$ momentum, per angular width $d\phi$, be denoted $\pe$.
Then in a similar spirit to eq. (\ref{zmom}), we have for the flux of the $\rho$
component of momentum
\beq
\label{rhomom}
\ps \sin \thetas \ = \ \parho(\thetas) \ + \ \pbrho(\thetas) \ + \ \pe \ ,
\eeq
where here
\beq
\parho(\thetas) \ = \ 2 \int_0^\thetas d\theta \sin^2 \theta \ = \ 
\thetas \ - \ \sin \thetas \cos \thetas 
\eeq
and
\beq
\pbrho(\thetas) \ = \  2 \int_\thetas^\pi d\theta \sin^2 \theta \ = \
\momrat \left ( \pi \ - \ \thetas \ + \ \sin \thetas
\cos \thetas \right ) \ .
\eeq
give the sources of $\rho$ momentum from the two winds respectively,
again integrated over the solid angle of embroiled gas appropriate for each wind.

It may at first glance
seem that the $\pe$ momentum flux in the $\rho$ direction appears somewhat
magically, but it can be physically traced to the curved axial symmetry, because
the azimuthal transport of vector momentum flux endemic to the high gas pressure
in an adiabatic shock will inevitably lead to an enhancement of
momentum flux in the $\rho$ direction.
This is similar to the way a billiard ball karoming around the
edges of a circular table exerts a continuous radially outward force.

Equations (\ref{zmom}) and (\ref{rhomom}) may be viewed as a matrix equation
in $(\momrat,\ps)$ in terms of the unknown $\pe$ and the given $\thetas$, 
which solves to
\beq
\label{momrat}
\momrat \ = \ {\sin \thetas \ - \ (\pe + \thetas) \cos \thetas \over
\sin \thetas \ + \ (\pi - \thetas ) \cos \thetas }
\eeq
and
\beq
\label{psfind}
\ps \ = \ {(\pi + \pe) \sin^2 \thetas \over \sin \thetas \ + \ (\pi - \thetas)
\cos \thetas} \ .
\eeq  
If we imagine that $\thetas$ is known from observation, then
the above represents two equations in the three unknowns $\momrat$, $\ps$, and $\pe$,
so solving them will require an auxiliary assumption about the energy transport.
We will address that assumption in the form of limiting cases that deal with the
degree of adiabaticity and the degree of wind mixing.

\section{The role of adiabaticity and mixing}

Our next requirement is to find appropriate constraints on the scalar 
energy flux along $\thetas$
to determine the appropriate value for $\pe$, which in turn gives us our fundamental
goal, $\momrat(\thetas)$, via eq. (\ref{momrat}).
We can do this most easily if we consider the limiting cases of radiative shocks,
adiabatic shocks with no mixing, and adiabatic shocks with complete mixing.
We shall see that this lists these effects in order of
increasingly explosive support of the bow-shock angle (and therefore in
descending order of 
the required value
of $\momrat$).

\subsection{Radiative shocks}

We first consider the simplest case of purely radiative shocks, for which gas pressure
plays no role and we may set $\pe = 0$.
This situation is handled in CRW; we include it here only for completeness.
The solution from eq. (\ref{momrat}) is immediately
\beq
\label{momratrad}
\momrat \ = \ {\tan \thetas \ - \ \thetas \over
\tan \thetas \ - \ \thetas \ + \pi } \ .
\eeq
Note that it is not necessary to make any assumptions about the degree of mixing in
the two shocked winds, nor to what extent
$\ps$ is locally
carried by a single speed or a spread in speeds within the shock,
because our result for $\thetas$
is a global attribute
of the momentum balance independently of how that momentum is partitioned
among shock components.
The resulting $\momrat(\thetas)$ is depicted in Fig. 1.

\subsection{Adiabatic shocks with no mixing}

We assume that
quasi-thin adiabatic shocks will ultimately convert all the heat
thermalized in the shock back into bulk kinetic energy flowing approximately along
$\thetas$, but the impact on $\ps$ will depend on whether the two
winds mix and reach a single combined velocity, or if they retain
their individual character and return only to their original terminal
speed, or some combination thereof.
First we treat the case where each wind returns adiabatically to its original terminal
speed without mixing.
This means that the scalar momentum flux missing from the two original winds
will all show up in the asymptotic flow along the shock angle, i.e., that
scalar momentum flux will all end up being $\ps$.
Hence instead of substituting for $\pe$ directly, we may replace eq. (\ref{rhomom})
with the constraint
\beq
\label{psmom}
\ps \ = \ 2\int_0^\thetas d\theta \ \sin \theta \ + \ 2 \momrat
\int_\thetas^\pi d\theta \ \sin \theta \ = \ 2 (1-\cos \thetas) \ + \
2 \momrat (1+\cos \thetas)  \ .
\eeq
This results in the solutions to eqs. (\ref{momrat}) and (\ref{psfind}) becoming
\beq
\label{momratadnomix}
\momrat \ = \ \tan^4 \left ( {\thetas \over 2} \right )
\eeq
and
\beq
\ps \ = \ 
2 (1-\cos \thetas) \ + \
2 \tan^4 \left ({\thetas \over 2} \right ) (1+\cos \thetas) \ ,
\eeq
subject to
\beq
\pe \ = \ {1 \over 8} \sec^4 \left ({\thetas \over 2 } \right ) [
8 \sin \thetas \ - \ \pi \cos (2\thetas) \ + \
4 (\pi-2\thetas) \cos \thetas \ - \ 3 \pi ] \ .
\eeq
The resulting $\momrat(\thetas)$ is depicted in Fig. 1.

\subsection{Adiabatic shocks with complete mixing}

If we make the opposite assumption that when the thermalized energy is
adiabatically returned to the flow along $\thetas$,
the winds are completely mixed and reach a single joint characteristic
velocity, then we get the maximal explosive support of the opening angle
$\thetas$.
The additional support comes from the fact that 
not only is the ram pressure perpendicular to the shock
thermalized, but even some of the ram pressure initally {\it along} the shock is thermalized,
when the winds mix and reach a common speed.
This is essentially an increase in gas pressure due to frictional heating, and so
we expect the smallest $\momrat(\thetas)$ in this case.
Here our global constraint on $\ps$ comes from the total scalar kinetic
energy flux and the total scalar mass flux
that enter the shock zone, which must in turn flow out at a single
characteristic speed, approximately along
$\thetas$, in a manner consistent with $\ps$.

Following this logic,
the scalar mass flux per wedge thickness $d\phi$, 
again in units where the total scalar momentum
flux from star A is $8\pi$, is
\beq
M_s \ = \ {2 \over \va} (1-\cos\thetas) \ + \ {2 \momrat \over \vb}
(1+\cos \thetas) \ ,
\eeq
and the scalar kinetic energy flux in those units is
\beq 
K_s \ = \ { \ps^2 \over 2 M_s} \ = \ 
(1 - \cos \thetas) \va \ + \ \momrat (1+\cos \thetas) \vb
\eeq
which results in the constraint
\beq
\ps \ = \ 2 \sqrt{ [1 \ - \ \cos \thetas \ + \ (1+\cos \thetas)
\momrat u] [1 \ - \ \cos \thetas \ + \ (1+\cos \thetas)
{\momrat \over u}] } \ ,
\eeq
where we have defined $u = \va/\vb$.
Applying eqs. (\ref{momrat}) and (\ref{psfind}) then gives 
\begin{eqnarray}
\label{momratadmix}
  \lefteqn{
\momrat \ = \ {\cos^2 \thetas \over 8 u (3 \cos \thetas \ - \ 1)} 
\sec^6 \left ({\thetas \over 2} \right )
[2(1+u^2)\cos^2 \thetas - 2(1+u^2)-u \sin^2 \thetas \tan^2 \thetas +} \nonumber \\
& & \sqrt{2} \sqrt{1+u+4u^2+u^3+u^4+(1-u)^2(1+u+u^2)\cos (2\thetas)} \sin
\thetas \tan \thetas ] \ ,
\end{eqnarray}
and the expressions for $\ps$ and $\pe$ may also be written in closed
form but they are quite long and involved.
Although it is not immediately obvious, the result in eq. (\ref{momratadmix})
does indeed reduce to eq. (\ref{momratadnomix}) when $u=1$, since then the presence or
absence of mixing is irrelevant to the global dynamics.
This result for $\momrat(\thetas)$ is also included in Fig. 1.

\section{Discussion}

Our fundamental result is that an increased flux of the momentum component away from
the axis is generated by the extreme heating of the shocked gas, and this
can substantially widen the asymptotic bow shock angle as seen in Fig. 1.
It is also clear that the bow shock angle reaches $90^\circ$ when $\momrat=1$
for any of the limits of radiative or adiabatic cooling, as would be expected
from symmetry requirements.
The figure shows that for asymptotic opening half-angles of roughly 50 degrees,
for example, adiabaticity roughly halves the weak-wind momentum
flux required to support that shock geometry, and if winds with a 
fairly extreme factor 4 contrast in
terminal speeds are mixed, it will halve the requirement yet again.
Indeed, the results from eqs. (\ref{momratrad}), (\ref{momratadnomix}), and
(\ref{momratadmix}) give that the required $\momrat$ for a 50 degree asymptotic
half-angle are 0.1, 0.05, and 0.027 respectively.
The physical source of these differences can be traced in a schematic
yet quantitative
way by examining scaling laws generated in the limit $\momrat \ll 1$, as
we analyze next.
\begin{figure}[t]
\begin{center}
\plotfiddle{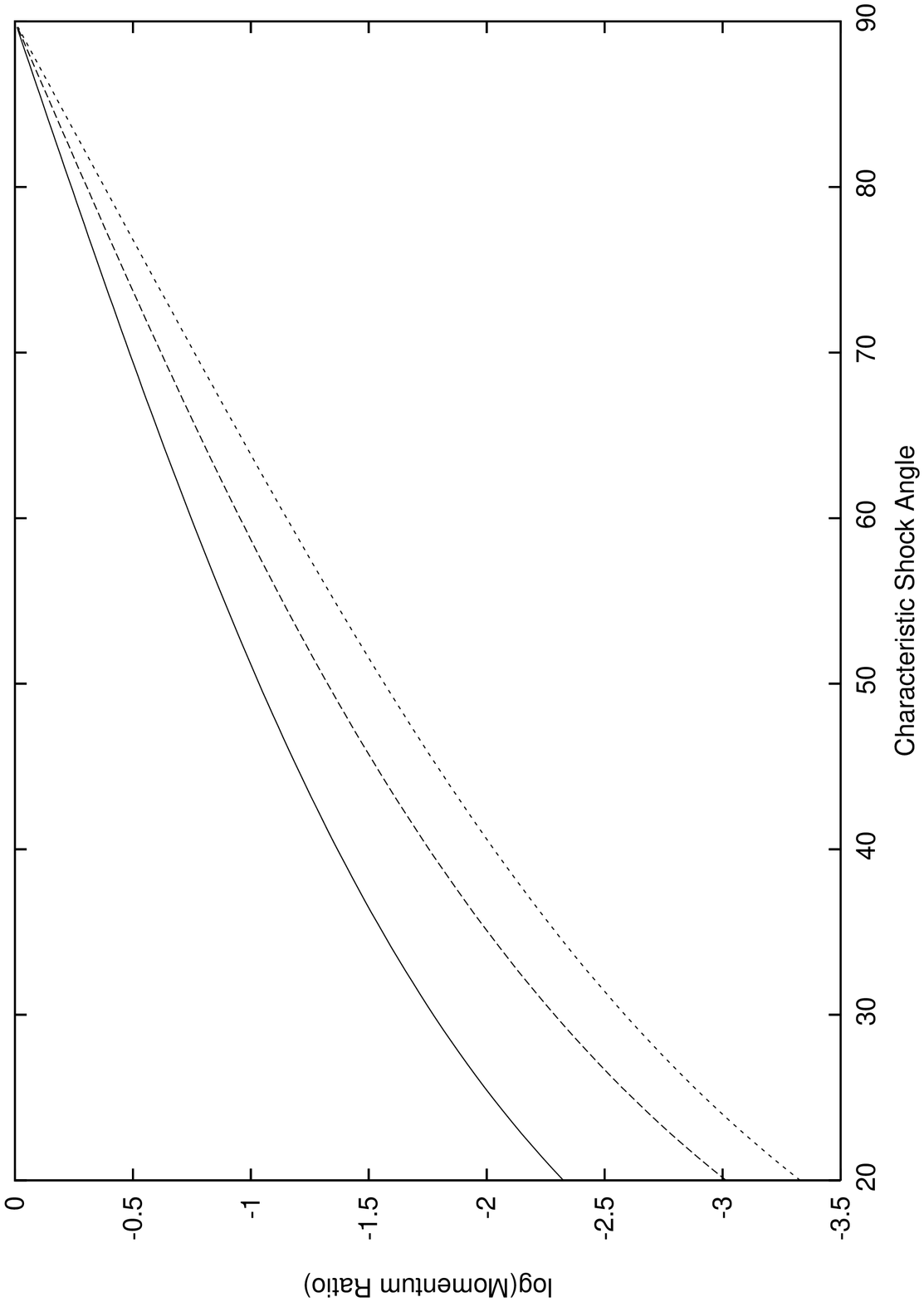}{2.6in}{-90.}{330.}{450.}{0}{0}
\caption{
The log (base 10) of the wind momentum ratio
$\momrat$ that would be necessary to sustain a characteristic shock angle $\thetas$
(in degrees),
for radiative shocks (solid curve), adiabatic shocks without mixing (dashed
curve), and adiabatic shocks with complete mixing and a terminal speed
contrast of $u = 4$ (dotted curve).  Note that the possibility of a direct
collision with the photosphere of the weak-wind star is not considered, even
for small $\momrat$.}
\end{center}
\end{figure}

\subsection{Winds with extremely low momentum-flux ratio}

The above results simplify in the limit $\thetas \ll 1$, which occurs
when $\momrat \ll 1$, so it is informative to consider what physical
insights may be conferred in that simple limit.
Note these limits apply asymptotically only when the weak-wind star
has a negligibly small radius; in real situations, when $\momrat$ is small enough
for asymptotic expressions to apply, the possibility must be considered
separately that the strong wind may
crash directly into the photosphere of the companion, invalidating our assumptions.
Nevertheless, the low-$\momrat$ limits do convey general insights into the reasons that 
different degrees of adiabaticity require different amounts of weak-wind momentum to support
the shock cone at a given opening angle $\thetas$, and it is one of the
primary advantages of closed-form expressions that they submit to this
type of scaling analysis.

We begin by noting that if one adopts the local approximation of equating
the perpendicular momentum fluxes across the shock front
(e.g., Dyson 1975; Luo et al. 1990; Stevens et al. 1992; Antokhin et al. 2004;
Falceta-Goncalves, Abraham, \& Jatenco-Pereira 2008) and 
extrapolates it globally to the asymptotic
opening angle, one should expect the required $\momrat$ to scale like $\thetas^2$ 
when $\thetas \ll 1$.
This is because the strong wind effectively has no inertia after it is shocked,
so the weak wind bears the full burden at every point along the shock
of maintaining that shock against the strong wind ram pressure.
The weaker wind can maintain
a perpendicular ram balance against at most a solid-angle fraction of order $\momrat$ of the
strong wind, and $\thetas^2$ determines the
solid-angle fraction subtended by the interaction zone, so $\momrat \propto \thetas^2$.

However, when a more accurate global accounting of the momentum requirements
is undertaken, in the limit of radiative shocks we find
from eq. (\ref{momratrad}) that
\beq
\momrat \ \cong \ {\thetas^3 \over 3 \pi} \ ,
\eeq 
as seen in the exact result of CRW and the heuristic fit of Eichler \& Usov (2003).
The extra power of $\thetas$ may be interpreted as being due to the fact
that the global requirement for the weak wind is not to {\it stop} the strong wind,
but merely to {\it deflect} it through an angle $\theta_s$.
So the $\momrat$ momentum flux must deflect through an angle $\thetas$ a fraction
$\sim \thetas^2$ of the strong wind, requiring $\momrat \sim \thetas^3$.
Indeed, we can be even more quantitative and note that the strong-wind momentum
flux along the $z$ direction is $2 \int_0^\thetas d\theta \ \sin \theta \cos \theta
\cong \thetas^2$, and the deflecting momentum flux in the $\rho$ direction 
from the weak wind is
$2\momrat\int_\thetas^\pi d\theta \sin^2 \theta \cong \pi \momrat$, along with
a contribution in the $\rho$ direction from the strong wind itself of
$2\int_0^\thetas d\theta \ \sin^2 \theta \cong 2\thetas^3/3$.
Adding the momentum flux in the $\rho$ direction to that in the $z$ direction bends
the total shocked momentum flux an angle $\thetas$, where from the above estimates we have 
\beq
\thetas \ \cong \ {\pi \momrat  \over \thetas^2} \ + \ {2 \over 3} {\thetas^3 \over
\thetas^2} \ ,
\eeq
which results directly in $\momrat \cong \thetas^3/3\pi$.

For adiabatic shocks with no mixing, our approximation in eq. (\ref{momratadnomix})
yields when $\thetas \ll 1$
\beq
\momrat \ \cong \ {\thetas^4 \over 16} \ .
\eeq
Here we find yet another added power of $\thetas$, this time because the explosive
heating of the shocked winds, upon re-expansion away from the axis, provides substantial
momentum support for the bending of the shock angle.
Indeed, the combination of the pre-shocked momentum flux in the $\rho$ direction, and the 
explosive gas pressure contribution in that direction, produce {\it almost}
enough momentum to support a small deflection $\thetas$ by themselves, without
help from the weak
wind.
Hence the weak wind needs to provide only a small ``coaxing'' on top of these two
momentum fluxes along $\rho$ (each which scales $\sim \thetas^3$), so this coaxing
appears at an even higher order of $\thetas$, at order $\thetas^4$.
So this analysis elucidates the physical reasons 
why radiative shocks require less weak-wind momentum flux
to maintain a given narrow shock angle than would a putative
perpendicular ram balance, and adiabatic shocks require less still.

Considering the case of adiabatic shocks with complete mixing, we find from
eq. (\ref{momratadmix}) that when $\thetas \ll 1$,
\beq
\momrat \ \cong \ {\thetas^4 u \over 8 (1+u^2)} \ ,
\eeq
where again $u$ is the ratio of the terminal speeds in the two winds, and it does
not matter which wind is in the numerator, only the contrast expressed by $u$.
Here we see the now-familiar $\theta^4$ scaling of adiabatic shocks,
but we also find that when there is a strong contrast in the wind speeds, adiabatic
mixing allows additional thermalization and additional explosive expansion away from
the axis, further supporting the deflection of the strong wind and allowing for
an even smaller $\momrat$ to suffice, in light of the identity $u/8(1+u^2) \ < \ 1/16$.

\section{Conclusions}

We use global momentum-flux considerations in
the context of a characteristic-angle shock approximation to
derive the resulting asymptotic opening angle of shocked gas for two colliding spherical
winds, for either fast or slow radiative cooling, with complete or limited
mixing.
Hence this may be viewed as an extension of the CRW approach to global
shock characteristics in the presence of significant adiabatic cooling.
For intermediate levels of adiabaticity and mixing,
informal interpolation of our results
would seem preferable to an effort to track the 
transitional physics in detail,
given the rough character
of the approximations used.

We find that adiabaticity measurably widens the asymptotic characteristic angle
of the wind interaction, or for a fixed observed
opening angle, 
significantly reduces the associated wind momentum-flux ratio that would support it,
as described in eqs. (\ref{momratrad}) and (\ref{momratadnomix}).
Mixing of winds with different terminal speeds would have no additional effect in the
radiative limit, but in the adiabatic limit further reduces the inferred wind
momentum-flux ratio, as seen in eq. (\ref{momratadmix}).
Hence, the observed angle, the wind momentum-flux ratio, and the degree of adiabaticity
and mixing, all form a set of parameters that permit knowledge about some
to be used to infer or constrain the others.
In particular, observations of the characteristic wind interaction angle, whether
from spectra or visible dust formation, can be used to draw inferences about the
character of the colliding winds.

The nature of the approximation is certainly highly idealized, as
radiative shocks are subject to shear instabilities, adiabatic shocks suffer
explosive spreading of the shocked gas, and mixing can result in a range of 
flow speeds instead of a single uniform one, so our approximation faces significant
limitations in practice.
Also, clumping in the wind might present additional challenges (although Pittard 2007
finds that clump winds collide in a broadly similar way to smooth winds).
Overall, the goal is to provide a straightforward way to obtain analytic closed-form
expressions which elucidate certain basic physical differences, which are intended
to inform a new vocabulary for unifying the discussion around hydrodynamic simulations
of wind/wind collisions over a wide range of circumstances.
Future hydrodynamical simulations will be needed to investigate the proper interpretation
of the concept of a characteristic quasi-thin shock angle, in the face of realistic
complications in that idealization.

For example, Pilyugin \& Usov (2007) find that equal-strength adiabatic winds 
collide in such a way as to generate so much pressure-driven expansion of the shock region
that ultimately both winds are entirely embroiled.
Nevertheless, we point out that
once the gas adiabatically cools and returns to some approximation of its original terminal
speed, a prevailing feature will be the contact discontinuity along the central
plane, and that is where our analysis would locate the ``characteristic angle''
of the interaction for equal winds.
So even when a thin shocked layer is not physically realized, there may yet be
value in thinking in terms of global momentum considerations and characteristic
interaction angles.

Thus our approximate results are here intended to provide a benchmark against which
to compare and interpret more detailed simulations, moreso than as a quantitatively
accurate description of the detailed nature of the wind interaction.
For highly unequal winds,
the most natural interpretation of the derived angle will be the
working surface in the stronger wind, but detailed simulations are needed to
verify this expectation.
Since the greater the adiabaticity, the wider the characteristic angle 
(especially for strong mixing),
these results are intended to help form expectations about the influences
of adiabaticity and mixing on any particular observed bow-shock geometry,
especially at early stages of the analysis when few constraints on the wind collision
may easily be determined.
Disentangling the independent parameters of mass and momentum fluxes in colliding winds
benefits from consideration of all possible diagnostic constraints, and the global
treatment here may be used to complement more detailed studies of the interaction
closer to the stagnation zone.

\blankline
\noindent{\bf References}
\blankline
\noindent
Antokhin, I. I., Owocki, S. P., \& Brown, J. C. 2004, ApJ, 611, 434

\noindent
Eichler, D. \& Usov, V. 1993, ApJ, 402, 271

\noindent
Canto, J., Raga, A. C., \& Wilkin, F. P. 1996, ApJ, 469, 729

\noindent
Comeron, F. \& Kaper, L. 1998, A\&A, 338, 273

\noindent
Dyson, J. 1975, Ap\&SS, 35, 299

\noindent
Falceta-Goncalves, D., Abraham, Z., \& Jatenco-Pereira, V. MNRAS, 2008, 383, 258

\noindent
Girard, T. \& Willson, L. A. 1987, A\&A, 183, 247

\noindent
Hill, G. M., Moffat, A. F. J., \& St-Louis, N. 2002, MNRAS, 335, 1069

\noindent
Ignace, R., Bessey, R., \& Price, C. S. 2009, MNRAS, 395, 962

\noindent
Lemaster, M. N., Stone, J., M., \& Gardiner, T. A. 2007, ApJ, 662, 582

\noindent
Luhrs, S. 1997, PASP, 109, 504

\noindent 
Luo, D., McCray, R. \& Mac Low, M.-M. 1990, ApJ, 362, 267

\noindent
Mac Low, M.-M., Van Buren, D., Wood, D. O. S., \& Churchwell, E. 1991, ApJ, 369, 395

\noindent
Pilyugin, N. N. \& Usov, V. V. 2007, ApJ, 655, 1002

\noindent
Pittard, J. 2007, ApJ, 660, L141

\noindent
Rauw, G., Crowther, P. A., De Becker, M., Gosset, E., Naze, Y., Sana, H.,
van der Hucht, K. A., Vreux, J.-M., \& Williams, P. M. 2005, A\&A, 432, 985

\noindent
Shore, S. N. \& Brown, D. N. 1988, ApJ, 334, 1021

\noindent
Stevens, I. R., Blondin, J. M., \& Pollock, A. M. T. 1992, ApJ, 386, 265

\noindent
St-Louis, N., Moffat, A. F. J., Marchenko, S., \& Pittard, J. 2005, ApJ, 628, 953

\noindent
St-Louis, N., Willis, A. J., \& Stevens, I. R., 1993, ApJ, 415, 298

\noindent
Tuthill, P. G., Monnier, J. D., Lawrance, N., Danchi, W. C., Owocki, S. P., \&
Gayley, K. G. 2008, ApJ, 675, 698

\noindent
Tuthill, P. G., Monnier, J. D., Tanner, A., Figer, D., Ghez, A., \& Danchi, W. 2006, Sci, 313, 935

\noindent
Wilkin, F. P. 1996, ApJ, 459, L31

\end{document}